\documentclass[aps,prl,reprint,twocolumn,floatfix]{revtex4-2}
\usepackage{graphicx}
\usepackage{dcolumn}
\usepackage{bm}
\usepackage[normalem]{ulem}
\usepackage{xcolor}
\usepackage{comment}
\usepackage{amsmath}
\usepackage{siunitx}
\usepackage{dsfont}
\usepackage{float}
\usepackage{placeins}
\usepackage[shortlabels]{enumitem}
\usepackage{soul}
\usepackage{booktabs}

\usepackage{hyperref}

\newcommand{\up}{\uparrow}
\newcommand{\dn}{\downarrow}
\DeclareSIUnit\angstrom{\text{Å}}

\usepackage{physics}
\begin{document}

\title{Local magnetic resonance of scalar spin chirality}

\author{Mar Ferri-Cortés$^{1,2}$, J. Fern\'{a}ndez-Rossier$^{2}$}

\affiliation{$^1$ Departamento de F\'isica Aplicada, Universidad de Alicante, 03690 San Vicente del Raspeig, Spain}
\affiliation{$^2$International Iberian Nanotechnology Laboratory (INL), Av. Mestre Jos\'e Veiga, 4715-330 Braga, Portugal }

\begin{abstract}

Quantum states with finite scalar spin chirality carry an orbital magnetic moment that has eluded direct measurement. We show that a local drive breaks the cyclic symmetry and activates chirality-changing transitions. We propose two complementary probes: direct chirality resonance and chirality-resolved spin resonance. Lindblad simulations demonstrate their feasibility in locally addressable spin platforms, including scanning tunneling microscopy, silicon donors, and quantum dots.
\end{abstract}

\maketitle


\textit{Introduction---}Magnetic resonance has been used to probe both electronic and nuclear spins for almost 90 years now~\cite{rabi38,bloch46,purcell46}.  Magnetic resonance methods  rely on the  combined action of  static and  dynamic magnetic fields that couple to the magnetic moment of electrons and nuclear spins.  Therefore, nuclear magnetic resonance  (NMR) and electron spin resonance (ESR)  can be used to determine the magnetic moment of elementary quantum particles, either electrons or nuclei,  even if in most instances the experiment is carried out with ensemble techniques.   

Analogously, both Ferromagnetic~\cite{griffiths46} and Antiferromagnetic~\cite{kittel51,Johnson56} Resonance (FMR \& AFMR) techniques are routinely used to probe the collective spin magnetization of magnetically ordered materials~\cite{kataev24}. Here we extend the concept of magnetic resonance to probe the  {\em orbital} magnetic moment associated with a class of  collective spin states 
that have a finite  scalar spin chirality (SSC). In its simplest form, the SSC   was introduced\cite{wen1989} for states describing three or more  coupled spins $S=1/2$:
\begin{equation}
    \hat{\chi}_{i,j,k} = 
\vec{S}_i\cdot\left(\vec{S}_j\times\vec{S}_k\right).
    \label{eq:chirality}
\end{equation}
where $\vec{S}_i \equiv \frac{1}{2}(\sigma^{x}_{i}, \sigma^{y}_{i}, \sigma^{z}_{i})$ are the spin-$\frac{1}{2}$ operators at site $i$ and ($\sigma^x$, $\sigma^y$, $\sigma^z$) are the Pauli matrices.  The eigenvalues of the SSC operator are $\chi=0,\pm \frac{\sqrt{3}}{{4}}$. 

The SSC operator has attracted interest on multiple fronts. The classical limit of the SSC   characterizes  non-co-planarity of broken-symmetry spin textures~\cite{fert17} and controls the  magnitude of the anomalous Hall effect~\cite{taguchi01,dos2016}. In correlated electrons, it relates to the 
orbital currents, magnetoelectric coupling~\cite{trif08,Bulaevskii08,kamiya12, hosokoshi26}, Raman scattering~\cite{shastry90}, and the response to circularly polarized light~\cite{kitamura17}. 
In quantum information,  the SCC relates to the commutator of permutation operators~\cite{subrahmanyam94}, 
the chiral doublet has been considered as a quantum bit~\cite{scarola04,cao08,georgeot10,hsieh2010}, the expectation value of the SSC operator~\cite{tsomokos08}   is a witness of genuine tripartite entanglement~\cite{dur00},   and can be detected in quantum computers using the cycle test~\cite{reascos23}.

It was realized\cite{wen1989} that the SSC commutes with the Heisenberg Hamiltonian of an $N=3$ ring of symmetrically coupled spins with exchange $J$. As a result, the spectrum of the Hamiltonian comes out with an additional degeneracy so that the two $S=1/2$ doublets are degenerate. 
The SSC eigenstates with finite chirality $\chi=\pm \frac{\sqrt{3}}{{4}}$ are  
\begin{eqnarray}\label{eq:chiralstates}
 \ket{0}   &=& \frac{1}{\sqrt{3}}\left(\ket{\uparrow\downarrow\downarrow} + \xi \ket{\downarrow\uparrow\downarrow} + \xi^2\ket{\downarrow\downarrow\uparrow} \right)  \notag \\ 
    \ket{1}&=& \frac{1}{\sqrt{3}}\left(\ket{\uparrow\downarrow\downarrow} + \xi^2\ket{\downarrow\uparrow\downarrow} + \xi\ket{\downarrow\downarrow\uparrow} \right) 
\end{eqnarray}
where $\xi=e^{2\pi i/3}$ represents clockwise/anticlockwise spin circulation. States $\ket{0},\ket{1}$ define the 
chiral doublet  with $S^z=-1/2$, and analogous definitions for states $\ket{2}$ and $\ket{3}$, with $S^z=1/2$ (more details in section
~\ref{ap:B} of the End Matter).

Starting from a half-filled Hubbard model in the strong-coupling limit $U\gg t$, where $t$ and $U$ are the hopping and on-site Coulomb repulsion respectively, it was shown\cite{sen95,scarola04,Bulaevskii08} that the resulting Heisenberg spin Hamiltonian acquires an additional term:
\begin{equation}
    H_{\chi}= \lambda B \cos\theta \; \hat{\chi}_{1,2,3}\equiv -\hat{\mu}_{\chi} B \cos\theta
 \label{eq:chiralcoupling}
\end{equation}
where 
 $   \lambda= \frac{24t^3}{U^2}\frac{eA}{\hbar}$ and $B \cos\theta$ is the component of the magnetic field perpendicular to the plane of the triangle. 
{Eq. ~\ref{eq:chiralcoupling} implies the existence of an orbital magnetic moment associated with the SSC operator. It is thus possible to define a  chiral magnetic moment operator $\hat{\mu}_{\chi}$, with  eigenvalues 
\begin{equation}
\pm \mu_{\chi}= 
\pm \frac{6\sqrt{3}t^3}{U^2}\frac{eA}{\hbar} 
    \label{eq:chiralmoment}
\end{equation}
 where $A$ is the area of the triangle. Equations ~\ref{eq:chiralcoupling} and ~\ref{eq:chiralmoment} show that the eigenstates of the spin-chirality operator carry two distinct  {\em magnetic moments}: the conventional spin magnetic moment, and an additional orbital magnetic moment arising from the {\em charge circulation} of electrons around the ring, and hence fundamentally different  from  the effective orbital magnetic moment of neutral collective modes such as magnons~\cite{neumann20,Tang26}. 
Table ~\ref{tb:table_chiralmagmom} provides estimated values of the magnitude of this orbital moment for several physical systems. In most cases, the orbital magnetic moment associated with scalar spin chirality exceeds by far the typical nuclear magnetic moment. However, to the best of our knowledge, it  has not yet been directly measured experimentally. Details of these estimates are shown in section ~\ref{ap:A} of the End Matter.

\begin{table}[h] 
\centering
\begin{tabular}{cccccc}
\toprule
System &  $\frac{|t|}{U}$ & t $(meV)$ & $L$ (nm)  & $\delta_{\chi}/h (GHz)$ & $\mu_{\chi}(\mu_B$)\\  
\midrule
Si:P~\cite{kiczynski22} & $0.136$ & $3.4$ & $10 $  & $20.8$ & $0.74$\\  
Q. dots~\cite{dehollain20} & $0.0034$& $0.01$ & $200$   & $0.015$ & $5.5\times 10^{-4}$\\   
NG~\cite{Jacob2022} & $0.1$& $100$ & $1$    & $3.3$ & $0.12 $\\  
\bottomrule
\end{tabular}
\caption{Order of magnitude estimates of magnetic chiral moment for some representative systems. $B=1T$. $L$ is the lateral dimension of the triangle. See section~\ref{ap:A} of the End Matter  for more details.} 
\label{tb:table_chiralmagmom}
\end{table}

The goal of this work is to propose an experimental method, feasible with the state-of-the-art, to detect the orbital magnetic moment associated with scalar spin chirality. Specifically, we propose to use   local spin resonance, as opposed to conventional electron spin resonance where  the magnetic field of the microwave would be the same amplitude on the 3 sites of the trimer, on account of the huge ratio of the microwave wavelength and the trimer dimensions. We show below that a local excitation, acting preferentially on one spin,   induces transitions both between states of the same and different spin chirality, measure their energy splitting and obtain the chiral magnetic moment. 
This local spin excitation can be implemented for instance in  ESR with a scanning tunneling microscope (ESR-STM)~\cite{baumann15,yang17,yang21}, as well as in spin-based quantum bit architectures, such as phosphorus donors in Silicon\cite{kiczynski22,mkadzik22}  and quantum dots~\cite{dehollain20}, where local operations over a single qubit have to be carried out, and therefore, selective excitation of individual spins  is required.


\textit{Model---}We consider an antiferromagnetic spin trimer, described with the following Hamiltonian:
\begin{equation}
    H_0 = J \sum_{i=1}^3 \vec{S}_i\cdot\vec{S}_{i+1} +g \mu_B \vec{B}\cdot\sum_i \vec{S}_i
+ H_{\chi} ,
\label{eq:ham}
\end{equation}
where the terms are ordered according to decreasing energy scale, and $\vec{S}_4\equiv \vec{S}_1$, $g=2$, and $\mu_B$ is the Bohr magneton. We take an antiferromagnetic coupling ($J>0$) and $J> g\mu_B B $ so that the two  $S=1/2$ doublets are well below in energy from the excited $S=3/2$ quartet. The Zeeman coupling splits the four low-energy $S=1/2$ states into 2 doublets ( ($\ket{0},\ket{1}$) and ($\ket{2},\ket{3}$)), with $ S^z_{\rm tot} =\pm \frac{1}{2}$, respectively. 
The remaining degeneracy within each spin sector is associated with the SSC degree of freedom. Including the coupling between the chirality-induced orbital magnetic moment and the magnetic field (Eq. ~\ref{eq:chiralcoupling}) splits each doublet. Since the spin magnetic moment is much larger than the chiral orbital moment, the low energy spectrum has two weakly split doublets, separated by the Zeeman energy (see Fig.~\ref{fig:Ediagram}).  
The energy spectrum of the low energy quartet can be written as:
\begin{align}
E_{0,1} &= -\tfrac{3J}{4} - \tfrac{\Delta_z}{2} \mp \tfrac{\delta_\chi}{2},\nonumber\\
E_{2,3} &= -\tfrac{3J}{4} + \tfrac{\Delta_z}{2} \mp \tfrac{\delta_\chi}{2},
\label{eq:spectrum}
\end{align}

where $\Delta_z=g\mu_B B$ is the Zeeman splitting and $\delta_\chi$ is the chiral  splitting given by 
\begin{equation}
    \delta_\chi=2 \mu_{\chi}B\cos\theta= 2 \mu_{\chi}B_z
    \label{eq:chiralsplitting}
\end{equation}

\begin{figure}[t]
    \centering
    \includegraphics[width=1.0\linewidth]{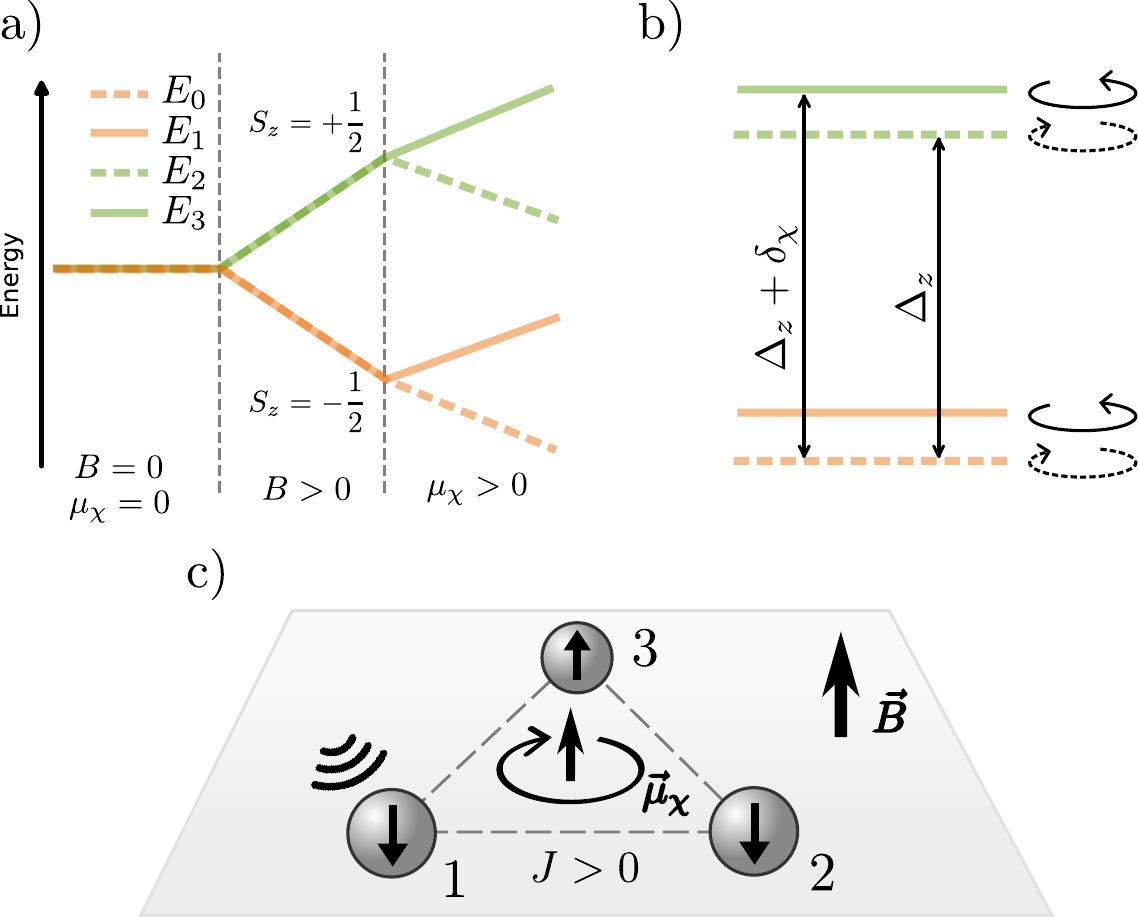}
    \caption{  Energy levels and physical system. a) Evolution of 4 lowest energy states of Hamiltonian \ref{eq:ham}. Colors denote spin sectors: orange and green for  $S=-\frac{1}{2}$ and $S=+\frac{1}{2}$ respectively. Line styles distinguish chiral sectors: dashed lines correspond to $-\chi$, while solid lines correspond to $+\chi$. b) Scheme denoting the two types of spin-flip transition, chirality preserving, and chirality flipping.  c) Schematic of the three-spin system on a surface, with off-plane  magnetic field $\vec{B}\parallel \hat{z}$ and a local AC driving on spin $1$.}
    \label{fig:Ediagram}
\end{figure}

\textit{Selection rules for chirality-resolved resonance---} The key ingredient of our proposal is that conventional and local 
spin resonance obey different symmetry selection rules. In conventional 
ESR, the driving field couples uniformly to all spins, and hence is proportional to $S_{\rm tot}^a$, with $a=x,y,z$.
Because the total spin operator is invariant under cyclic permutations 
of the sites, it commutes with the scalar spin chirality operator,
$\left[S_{\rm tot}^a,\hat{\chi}_{123}\right]=0$. Conventional ESR 
therefore preserves the chirality quantum number and cannot induce 
transitions between states of opposite chirality.

In contrast, a local drive acting on a single spin,
\begin{equation}\label{eq:driving}
    V(t)= g\mu_B b(t) 
    \left(\cos\alpha\, S_1^x + \sin\alpha\, S_1^z\right),
\end{equation}
breaks the $C_3$ rotational symmetry of the trimer and does not commute with the chirality operator because $\left[S_1^{a},\hat{\chi}_{123}\right]\neq 0$ for $a=x,y,z$.
In Eq.~\ref{eq:driving},  $\alpha$ controls the direction of the local perturbation in the $xz$ plane.
Without loss of generality, we label the driven  spin as site 1 (see Fig.~\ref{fig:Ediagram}c). Local excitation can be achieved by dynamical modulation of the exchange interaction with a 
spin-polarized tip~\cite{lado17}, or, for donors in silicon, by 
spectrally addressing individual spins via local hyperfine and Stark 
shifts~\cite{pla12,veldhorst15}.

Projected onto the low-energy basis, the local spin operators 
take the form
\begin{equation} \label{eq:spinproj}
S_1^z = \sigma_z \left(\tfrac{1}{6}\tau_0 - \tfrac{1}{3}\tau_x\right),
\quad
S_1^x = \sigma_x \left(-\tfrac{1}{6}\tau_0 + \tfrac{1}{3}\tau_x\right),
\end{equation}
where $\sigma_a$ and $\tau_a$ are Pauli matrices in the spin and chirality subspaces respectively. 
Both projected operators contain the $\tau_x$ matrix: the perturbation couples states of opposite chirality. Additionally, $S_1^x$ is off-diagonal in $\sigma$ and  therefore also flips the spin, while $S_1^z$ does not. Similar symmetry-breaking transitions may also arise from modulations of the Heisenberg exchange couplings  that also break $C_3$ symmetry\cite{cao08,trif08,hsieh2010}.  The advantage of the local spin operator is that it provides a direct and experimentally accessible route to induce and probe chirality-resolved spin resonances.

In the following,  we propose two different experimental procedures to measure the chiral magnetic moment using local spin resonance.
The first uses driving frequencies close to 
$\omega\simeq \delta_{\chi}/\hbar$ and reads the chiral splitting out directly through phase-sensitive homodyne detection within a single Zeeman sector; the second uses $\omega \simeq \Delta_z/\hbar$ and makes it possible to read out 
$\delta_{\chi}$ from the splitting of the spin-flip line into chirality-preserving and chirality-flipping components. We model 
both with a Lindblad master equation for the low-energy manifold in the rotating-wave approximation (RWA)\cite{Breuer2002}. Each simulation considers four dissipation parameters, the energy relaxation lifetimes $T_{1,s},T_{1,\chi}$  and the  pure dephasing rates $\gamma_{\phi,s},\gamma_{\phi,\chi}$, acting on the  chirality and spin channels. The coherence decay rate is given by $\frac{1}{T_{2,\chi/s}}\equiv \frac{1}{2T_{1,\chi/s}}+\gamma_{\phi, \chi/s}$ (see  End Matter ~\ref{ap:C}).

\textit{Direct chirality resonance---}
This approach relies on the fact that a spin-conserving  local driving ($\cos\alpha = 0$ in Eq.~(\ref{eq:driving}) can induce steady state coherence in the chiral channel, $\langle \tau_x (t)\rangle\neq 0$. By virtue of Eq.(~\ref{eq:spinproj}), 
this  induces a modulation of the longitudinal local magnetization (while the total magnetization is preserved) that can be probed using, for instance, magnetoresistive homodyne detection\cite{bae18,chen23,bae18}.   Since, global $S_z$ is conserved with $\cos\alpha=0$,  the two spin doublets decouple in two separated two-level systems. The low energy one, relevant for this experiment is given by: 
\begin{equation}\label{eq:Hchiral2level}
  H_{\text{eff}}(t) = \frac{\delta_\chi}{2}\,\tau_z + \hbar\Omega_R\cos(\omega t)\,\tau_x\,,
\end{equation}
with $\Omega_R = \Omega/3 =  g\mu_B b_0/(3\hbar)$.   The operator that drives this transition is also the one that reads it out: within each Zeeman sector, $S_1^z \propto \tau_x$ (Eq.~\eqref{eq:spinproj}). This dual role has a direct consequence for detection. 
In the rotating frame, the steady state solution of  this dissipative driven two level system is the standard Bloch solution\cite{bloch46}.
\begin{equation}\label{eq:XY_app}
  \langle \tau_x\rangle=
  \frac{T_{2,\chi}^2\,(\frac{\delta_\chi}{\hbar} - \omega)\,\Omega_R\,Z_0}{ D}
  \,,\qquad \langle \tau_y\rangle  = -\frac{T_{2,\chi}\,\Omega_R\,Z_0}{D}\,,
\end{equation}
where  $Z_0 = \langle\tau_z\rangle_{\text{eq}}$ is the thermal chirality polarization,   $D=1 + T_{2,\chi}^2(\frac{\delta_\chi}{\hbar} - \omega)^2 + T_{1,\chi} T_{2,\chi}\Omega_R^2$, 
and $T_{(1,2),\chi}$ are the chirality-relaxation and chirality-coherence times. In the numerical results of Fig.~\ref{fig:direct_resonance} we have assumed $k_BT= \frac{\delta_{\chi}}{4}$, so that $\langle Z_0 \rangle=-0.96$. 

\begin{figure}[t!]
    \centering
    \includegraphics[scale=0.45]{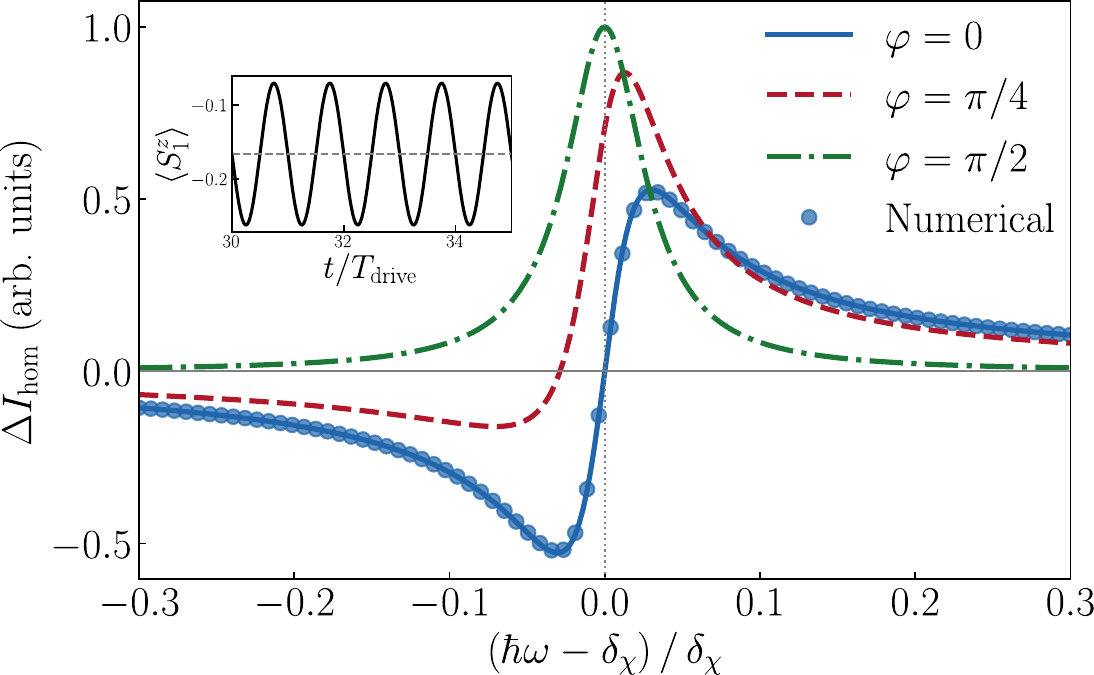}
    \caption{Direct chirality resonance via homodyne detection. Main panel: steady-state homodyne signal $\Delta I_{\rm hom}$ as a function of the normalised detuning $(\hbar\omega - \delta_\chi)/\delta_\chi$ for three values of the phase difference $\varphi$ (Eq.~\eqref{eq:Ihom}). Blue dots are the full four-level Lindblad steady-state simulation results, in agreement with   analytical theory (Eq.~\eqref{eq:XY_app}). Inset:  steady state local magnetization $\langle S_1^z \rangle$ at resonance ($\omega = \delta_\chi/\hbar$), in agreement with Eq.~\eqref{eq:S1z}.  Dashed grey line marks the equilibrium value of $-1/6$. Parameters:
    $B =1T$, 
$\mu_\chi=0.12   \mu_B$,  
    $\hbar\Omega = 200\,\text{neV}$, 
    $\Omega T_{1,\chi} =10$ , 
    $\gamma_{\phi,\chi}= \Omega/10$,
    $k_BT = \delta_\chi/4$ .}
    \label{fig:direct_resonance}
\end{figure}

In the lab frame, we have
\begin{equation}\label{eq:S1z}
  \langle S_1^z(t)\rangle = -\tfrac{1}{6} + \tfrac{1}{3}[\langle \tau_x \rangle \cos\omega t - \langle \tau_y \rangle \sin\omega t]\,,
\end{equation} 
in  agreement with our numerical solution of the Lindblad equation,  shown in the inset of Fig.~\ref{fig:direct_resonance}. Thus, 
we predict an  oscillation around the unperturbed value of the local magnetization, $-1/6$. 
Although the driving  leaves no imprint on the local DC magnetization,   the AC part can be probed with a 
phase-sensitive detection, such as the one used in ESR-STM, where 
homodyne detection provides the readout mechanism. Following the well-established treatment of tunneling magnetoresistance with an oscillating bias~\cite{bae18,chen23}, the time-averaged tunneling current acquires a resonant contribution (End Matter ~\ref{ap:D})
\begin{equation}\label{eq:Ihom}
  \Delta I_{\text{hom}} \propto
    \langle \tau_x \rangle \cos\varphi + \langle \tau_y\rangle\sin\varphi\,,
\end{equation}
where $\varphi$ is the phase lag between the RF bias that modulates the junction and the AC field that drives the
spin~\cite{chen23}. In our numerical simulations we take $\varphi=0$. The phase $\varphi$ determines the observed lineshape, as shown in Fig.~\ref{fig:direct_resonance}. The $\varphi = 0$ curve (solid blue) yields a dispersive (antisymmetric) lineshape whose zero-crossing locates $\delta_\chi$ independently of the linewidth; $\varphi = \pi/2$ (dash-dotted green) gives a purely absorptive Lorentzian peak; intermediate values produce Fano-like asymmetries ($\varphi = \pi/4$, dashed red). Regardless of the value of $\varphi$, the curve $\Delta I_{\text{hom}}(\omega)$ will display a  feature for $\hbar\omega=\delta_{\chi}$, that can be used to estimate this quantity.

\textit{Chirality-resolved spin resonance---}
We now consider a  complementary route to measure $\delta_\chi$, exploiting the spin-flip transitions near the Zeeman frequency. We assume  $k_BT \ll \Delta_z$, so that only the lowest two energy states, $|0\rangle$, $|1\rangle$ with $S_z = -1/2$ are occupied before the driving is turned on. The population of each of these states  depends on $\frac{k_BT}{\delta_{\chi}}$.  A weak AC field $b(t) = b_0\cos\omega t$ acting locally on spin~1, with frequency $\omega \simeq \Delta_z/\hbar$ induces transitions to the $S_z = +1/2$ manifold, as long as $\cos\alpha\neq 0$, because the transverse component $S_1^x$ drives spin-flip transitions (through $\sigma_x$). 

As a result, the ESR spectrum exhibits two classes of transitions, chirality preserving and chirality flipping, with frequencies 
 $\frac{\Delta_z}{\hbar}$ and 
   $\frac{\Delta_z\pm \delta_{\chi}}{\hbar}$ respectively.  
Consequently, the splitting of these resonance peaks provides direct access to $\delta_{\chi}$, and $\frac{d\delta_{\chi}}{dB_z}$
yields the chiral magnetic moment via Eq.~\eqref{eq:chiralsplitting}.

\begin{figure}[t!]
    \centering
    \includegraphics[scale=0.42]{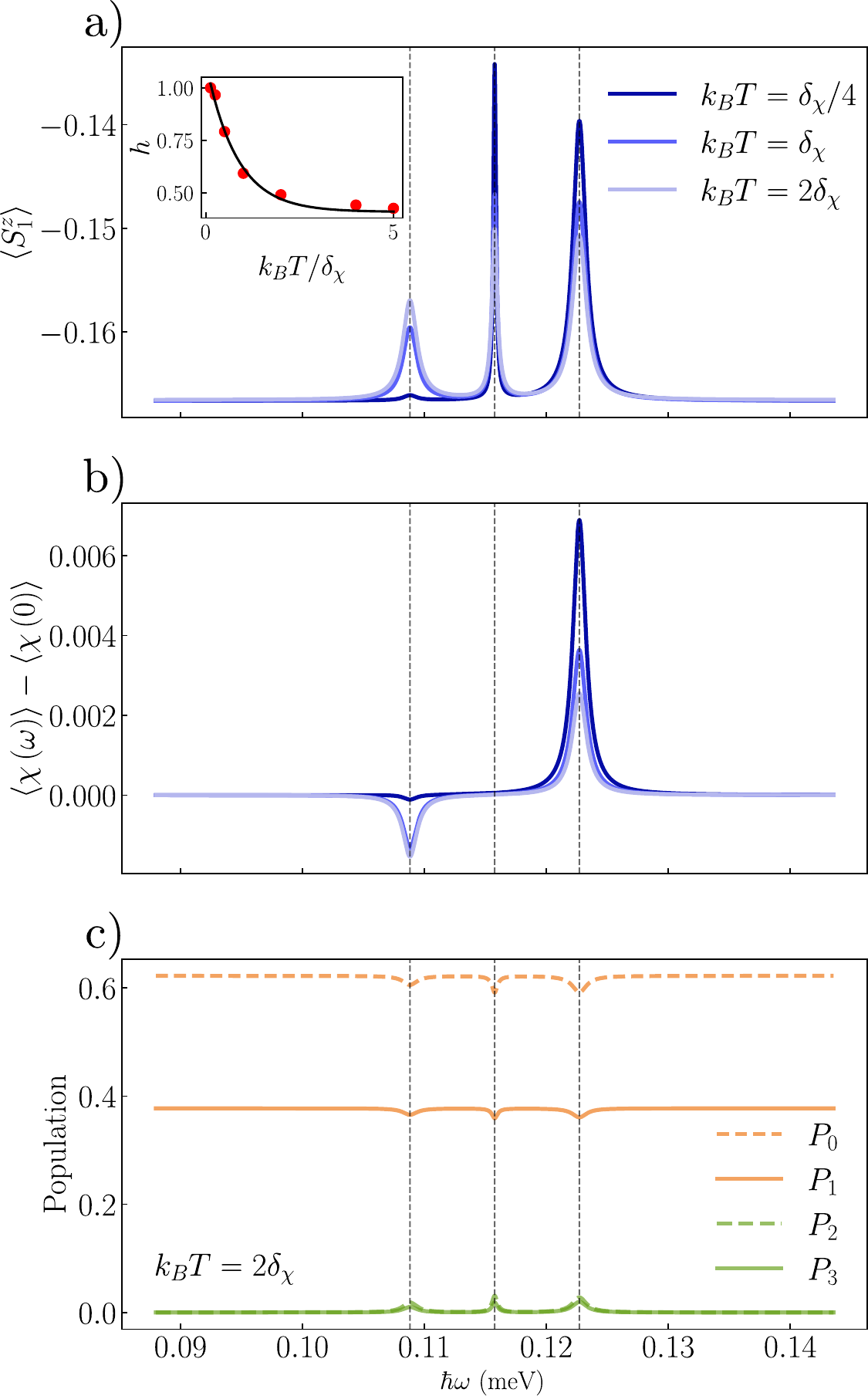}
    \caption{Simulation of the ESR-STM resonance. a)  $\langle S_1^z\rangle(\omega)$ for different temperatures, featuring  resonance peaks at $\hbar\omega =\Delta_z$ and $\hbar\omega = \Delta_z  \pm \delta_{\chi}$.The height of the  highest-energy   peak ($\hbar\omega = \Delta_z  + \delta_{\chi}$) is controlled by equilibrium occupation of the state $\ket{1}$, as shown in inset, where
    $h=\left|\frac{\langle S_1^z\rangle_{\rm res}- \langle S_1^z\rangle_{0}}{\langle S_1^z\rangle_{0}}\right|$. b) Corresponding $\langle \chi\rangle(\omega)$ curves  for the same set of temperatures. Expectedly, resonance peaks are only seen at the chirality flipping frequencies.  c) Corresponding populations of the low energy states for the high temperature example, $k_{B}T = 2 \delta_{\chi}$.  The parameters for the simulations are $B=1T$, $\hbar\Omega =200$neV, $\Omega T_{1,s}= 10$, $\gamma_{\phi, s} = \Omega/2$, $\Omega T_{1,\chi} = 1$ and $\gamma_{\phi, \chi} = 2\Omega$. The direction of the AC magnetic field is set with an angle $\alpha = \frac{\pi}{4}$. The values for $\hbar\omega$ of $\Delta_z$, $\Delta_z - \delta_{\chi}$ and $\Delta_z + \delta_{\chi}$ are shown in grey dashed lines. }
    \label{fig:ESR_STM}
\end{figure}

Experimentally, the ESR signal is obtained from the frequency-dependent change in the time-averaged local magnetization $\langle S_1^z \rangle$ via a single-spin observable, as realized experimentally in platforms such as ESR-STM~\cite{baumann15} or spin-to-charge conversion techniques~\cite{pla12}, e.g., in donor-based silicon qubits. This quantity exhibits resonant features when the driving frequency matches energy differences between eigenstates. Equivalent information is obtained from the steady-state populations, as resonant driving induces redistribution of occupation among eigenstates.


These results are summarized in Fig.~\ref{fig:ESR_STM}. The resonance peaks in $\langle S_1^z\rangle$ appear at the expected frequencies $\Delta_z/\hbar$ and $(\Delta_z\pm \delta_{\chi})/\hbar$. Their relative intensities depend on temperature. In the low-temperature limit, when $k_{B}T\ll \delta_{\chi}$, only the ground state $\ket{0}$ is populated, and the transition at $(\Delta_z -  \delta_{\chi})/\hbar$ is suppressed. As the temperature increases to $k_{B}T\sim \delta_{\chi}$, the population of $\ket{1}$ (see Fig.~\ref{fig:ESR_STM}c) enables this transition and produces an additional peak. The angle $\alpha$ controls the relative weight of the central peak and the chiral satellite peaks; we use $\alpha=\pi/4$ in Fig.\ref{fig:ESR_STM}.

Consistent features are also observed in the expectation value of the chirality. In this case, only transitions that change chirality contribute, resulting in peaks at $(\Delta_z\pm \delta_{\chi})/\hbar$, while the transition $|0\rangle\rightarrow|2\rangle$ is inside the same chirality sector and therefore remains invisible.

The resolution of a given peak  in a resonance experiment requires its linewidth to be smaller than the energy splitting to the neighboring transitions, i.e.
 $\delta_{\rm FWHM} \lesssim \delta_{\chi}$. For a conventional $S=1/2$ system\cite{bloch46}, at fixed Rabi coupling $\Omega$, the linewidth, given by the full width at half maximum (FWHM) of the resonance is set by the spin relaxation and spin decoherence times, 
 \begin{equation}
     \delta_{\rm FWHM}^s = \frac{2\hbar}{T_{2,s}}\sqrt{1 + \Omega^2T_{1,s}T_{2,s}}.
     \label{eq:FWHM}
 \end{equation}
Our system, however, is an effective 4-level system and we have to address the impact of the  chiral relaxation times, $T_{1,\chi}$ and $T_{2,\chi}$, on the line-widths of the three resonances. These transitions sit at energies $\Delta_z + b\delta_\chi$ ($b=0,\pm$) 
For that matter, we fix $T_{1,s}, T_{2,s}$ and we ramp the ratio $r=\frac{T_{1,\chi}}{T_{1,s}}=\frac{T_{2,\chi}}{T_{2,s}}$. Our numerical calculations (see Fig. \ref{fig:4}a)  show two types of behaviour. The linewidth of the  chirality conserving transition, $b=0$, departs from the TLS prediction of Eq.~\ref{eq:FWHM} for small $r$ and approaches the pure-spin value asymptotically as $r$ grows. In contrast, the linewidth of the chirality flipping transitions, $b=\pm$, are always clearly larger than the spin-conserving transition, and is the same for both transitions.  We also compute the height of the three transitions as a function of the ratio $r$ (see Fig. \ref{fig:4}b).  Our results can be rationalized in terms of both the thermal population of the initial states, and the evolution of the line-widths . Importantly,  we find that, in all the range $10^{-1}\leq  r\leq 10^{2}$, the predicted line-widths remain much smaller than the chiral splitting for the case where the magnetic chiral moment is 0.12 $\mu_B$, ensuring its visibility in local spin resonance experiments. 



\begin{figure}[t!]
    \centering
    \includegraphics[scale=0.26]{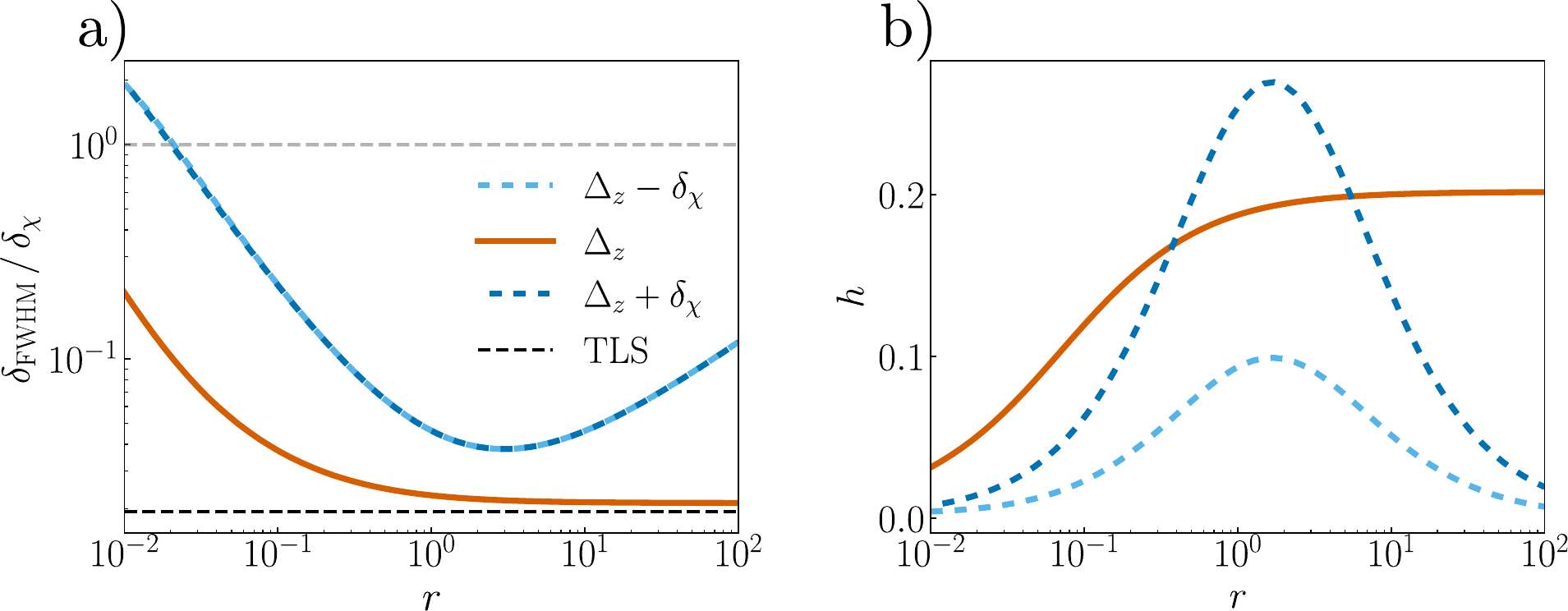}
    \caption{Analysis of the resolution in the resonance experiment. a) The value of $\delta_{\rm FWHM}$ normalized to $\delta_{\chi}$ as a function of the ratio $r=\frac{T_{1,\chi}}{T_{1,s}}=\frac{T_2^\chi}{T_{2,s}}$. The grey dotted line shows the visibility limit, where the width of the peaks is equal to the energy difference with other transitions. The black dashed line shows the analytical two-level system (TLS) value from Eq.~\ref{eq:FWHM} , which is shown to be a threshold. b) The normalized height of the peaks, $h=\left|\frac{\langle S_1^z\rangle_{\rm res}- \langle S_1^z\rangle_{0}}{\langle S_1^z\rangle_{0}}\right|$, as a function of the ratio. The parameters used in this simulation are the same as in Fig.~\ref{fig:ESR_STM}}.
    \label{fig:4}
\end{figure}

\paragraph{Discussion and Conclusions.---}

Quantum states with finite scalar spin chirality~\cite{wen1989} and their associated orbital magnetic moment~\cite{sen95} have been studied theoretically for more than three decades~\cite{trif08,Bulaevskii08,kamiya12,hosokoshi26,shastry90,kitamura17,subrahmanyam94,scarola04,cao08,
georgeot10,hsieh2010,tsomokos08,reascos23}, yet a direct spectroscopic determination of the orbital magnetic moment associated with a quantum chirality doublet remains lacking, to our knowledge.
Here we have shown that local spin resonance can overcome the selection rules of uniform driving and provide direct access to the chiral splitting $\delta_\chi$  and hence to the orbital moment $\mu_\chi$, through two complementary protocols. First, a local longitudinal drive induces transitions between opposite chiralities within a fixed $S_z$
 manifold, generating an oscillating local magnetization that can be detected by phase-sensitive homodyne detection.  Second, local spin-flip resonance produces chirality-changing satellite transitions at $\Delta_z\pm\delta_\chi$
 around the chirality-preserving Zeeman resonance at $\Delta_z$.
Our Lindblad simulations show that these signatures remain resolvable over a broad range of relaxation and decoherence parameters. These results establish local magnetic resonance as a route to measure the orbital magnetic moment associated with scalar spin chirality. Specifically,  spin qubit systems such as Phosphorous donors in Silicon\cite{kiczynski22} and semiconductor quantum dots \cite{dehollain20}, as well as spin-on surfaces probed with ESR-STM\cite{baumann15} provide particularly promising platforms to carry out the proposed experiments.

\vspace{1cm}

\paragraph{Acknowledgments.---}

We acknowledge J. C. Henriques for useful comments.  
M.F.C.
 acknowledges funding from Generalitat Valenciana
(CIACIF/2021/434).
J.F.-R.  
acknowledges financial support from 
 SNF Sinergia (Grant Pimag),
J.F.-R. and M.F.C.  acknowledge funding from
%
Generalitat Valenciana (Prometeo2021/017
and MFA/2022/045)
and
MICIN-Spain (Grants No. PID2022-141712NB-C22 and PRTR-C17.I1) 
We acknowledge the use of OpenAI's ChatGPT as a research aid for theoretical discussions, numerical and coding tasks, and manuscript editing. We have carefully revised all the results. Accordingly, we take full responsibility of the presented material.

\bibliography{biblio}

%

\onecolumngrid
\vspace{0.5em}
\begin{center}
  \textbf{\large End Matter}
\end{center}
\vspace{0.3em}
\twocolumngrid

\setcounter{equation}{0}
\renewcommand{\theequation}{EM\arabic{equation}}

\newcounter{appsec}
\renewcommand{\theappsec}{\Alph{appsec}}
\newcommand{\AppSec}[2]{%
  \refstepcounter{appsec}%
  \section*{\theappsec. #2}%
  \label{#1}%
}

\AppSec{ap:A}{Chiral orbital magnetic moment from a Hubbard model}

To estimate the chiral orbital magnetic moment $\mu_\chi$ entering Eq.~\ref{eq:chiralcoupling} of the main text, we start from a single-band Hubbard model on a triangular ring with $N$ sites at half filling and consider its response to a magnetic flux through the ring:
\begin{equation}
H = -\sum_{i,\sigma} t(\phi)\,c^\dagger_{i,\sigma}c_{i+1,\sigma} + {\rm h.c.} + U\sum_i n_{i,\up}n_{i,\dn}.
\end{equation}
The flux per bond enters through the Peierls phase
\begin{equation}
t(\phi) = t\,e^{i\phi},\qquad
\phi = \frac{2\pi}{N}\,\frac{\Phi_B}{\Phi_0},
\end{equation}
where $\Phi_B = B_zA$ is the magnetic flux threading the loop, $A$ is the area, and $\Phi_0 = h/e$ is the flux quantum. The total phase around the loop is $\Phi = N\phi$. In the following we assume $N=3$

The orbital magnetic moment of an eigenstate with energy $E_n$ is
\begin{equation} \label{eq:mudef}
\mu_n \equiv -\frac{\partial E_n}{\partial B}
= -\frac{\partial E_n}{\partial \Phi}\,\frac{2\pi A}{\Phi_0}.
\end{equation}

%

In the strong-coupling regime $U\gg t$, the half-filled Hubbard model
maps onto a Heisenberg spin model with additional chiral terms.
Parametrising the flux derivative as
\begin{equation}
\frac{\partial E_n}{\partial \Phi} = \alpha_n\,t\,\left(\frac{t}{U}\right)^2,
\end{equation}
where the sign in \ref{eq:mudef} is absorbed in $\alpha_n$, which is an order-unity coefficient. 
Therefore, for  we obtain
\begin{equation} 
\mu_n = \alpha_n\,t\,\left(\frac{t}{U}\right)^2 \,\frac{2\pi A}{\Phi_0}.
\end{equation}
Taking into account the Bohr magneton is $\mu_B = \frac{\hbar e}{2m_e}$, the total chiral magnetic moment of the loop in $\mu_B$ units is:
\begin{equation}
    \frac{\mu_n}{\mu_B} = 2\alpha_n  \frac{m_e A t^3}{\hbar^2 U^2},
\end{equation}
where the area of an equilateral triangle is $A= \frac{\sqrt{3}}{4}L^2$, where $L$ is the side length. For the triangle, the numerical evaluation yields $\alpha_n = \pm 6\sqrt 3$ (independent of $t/U$), recovering the chiral moment defined in Eq.~\ref{eq:chiralmoment}. The values in Table~\ref{tb:table_chiralmagmom} follow by inserting platform-specific $t$, $U$ and $L$.

\AppSec{ap:B}{Projection of the local spin operators
onto the chiral doublet manifold}

The selection rules for the local resonance scheme follow from the
form of the single-site operators $S_1^z$ and $S_1^x$ once they are
restricted to the four-dimensional low-energy manifold spanned by the
chiral states $|0\rangle,|1\rangle,|2\rangle,|3\rangle$ defined in
Eq.~\ref{eq:chiralstates} of the main text. As we have $J\gg\Delta_z$, we discard the  $S=3/2$ quartet.
 A direct evaluation of the matrix elements
in the chiral basis, using the identity $1+\omega+\omega^{2}=0$ and
the action of $S_1^z$ on the product states $|{\up\dn\dn}\rangle$,
$|{\dn\up\dn}\rangle$, $|{\dn\dn\up}\rangle$ that span the
$S^z_{\rm tot}=-1/2$ sector (and their spin-flipped counterparts in
the $S^z_{\rm tot}=+1/2$ sector), yields
\begin{equation}
S_1^z =
\begin{pmatrix}
-\tfrac{1}{6} & +\tfrac{1}{3} & 0 & 0 \\[2pt]
+\tfrac{1}{3} & -\tfrac{1}{6} & 0 & 0 \\[2pt]
0 & 0 & +\tfrac{1}{6} & -\tfrac{1}{3} \\[2pt]
0 & 0 & -\tfrac{1}{3} & +\tfrac{1}{6}
\end{pmatrix}
= \sigma_z\!\left(\tfrac{1}{6}\,\tau_0
- \tfrac{1}{3}\,\tau_x\right),
\label{eq:Sz1_matrix}
\end{equation}
and
\begin{equation}
S_1^x =
\begin{pmatrix}
0 & 0 & -\tfrac{1}{6} & +\tfrac{1}{3} \\[2pt]
0 & 0 & +\tfrac{1}{3} & -\tfrac{1}{6} \\[2pt]
-\tfrac{1}{6} & +\tfrac{1}{3} & 0 & 0 \\[2pt]
+\tfrac{1}{3} & -\tfrac{1}{6} & 0 & 0
\end{pmatrix}
= \sigma_x\!\left(-\tfrac{1}{6}\,\tau_0
+ \tfrac{1}{3}\,\tau_x\right),
\label{eq:Sx1_matrix}
\end{equation}
where $\sigma_a$ and $\tau_a$ ($a=0,x,y,z$) denote Pauli matrices
acting in the spin and chirality subspaces, respectively, with
$\sigma_0$ and $\tau_0$ the corresponding identity operators. Here
$\sigma_z = \pm 1$ for $S^z_{\rm tot} = \pm 1/2$, so that in the
basis ordering used above $\sigma_z = {\rm diag}(-1,-1,+1,+1)$.

The opposite signs of the diagonal entries in the two Zeeman blocks
of $S_1^z$ are required by the global spin-flip symmetry of the
trimer, which sends $S_1^z \to -S_1^z$ together with $S^z_{\rm tot}
\to -S^z_{\rm tot}$, and equivalently by the sum rule
$\sum_i\langle S_i^z\rangle = S^z_{\rm tot}$ combined with the
$C_3$ equivalence of the three sites, which gives
$\langle S_1^z\rangle = S^z_{\rm tot}/3 = \mp 1/6$ in the lower and
upper Zeeman sectors, respectively.

Both projected operators contain the  $\tau_x$ matrix, which couples
states of opposite chirality. The $\sigma_x$ factor in $S_1^x$ flips,
in addition, the total spin. The perturbation therefore acts on 
the unperturbed four-level Hamiltonian
\begin{equation}
    {\cal H}_0 \equiv \frac{\Delta_z}{2}\sigma_z + 
    \frac{\delta_\chi}{2}\tau_z
\end{equation}
as
\begin{equation}
    {\cal V}(t) = g\mu_B b(t)\,
    \left(-\tfrac{1}{6}\tau_0 + \tfrac{1}{3}\tau_x\right)
    \left(\cos\alpha\,\sigma_x - \sin\alpha\,\sigma_z\right).
\end{equation} 
This perturbation activates therefore three classes of transitions
in the low-energy manifold: chirality-preserving spin flips at
frequency $\Delta_z/\hbar$ (from the $\sigma_x\,\tau_0$ piece of
$S_1^x$); chirality-flipping spin flips at frequencies
$(\Delta_z\pm\delta_\chi)/\hbar$ (from $\sigma_x\,\tau_x$); and
chirality-flipping intra-Zeeman transitions at frequency
$\delta_\chi/\hbar$ (from the $\sigma_z\,\tau_x$ piece of $S_1^z$). These are the transitions exploited in the direct chirality resonance of Fig.~\ref{fig:direct_resonance}, while the two satellite peaks at $\omega=(\Delta_z\pm\delta_\chi)/\hbar$ are the fingerprint of the chiral orbital magnetic moment in the spin-flip window of Fig.~\ref{fig:ESR_STM}.

\AppSec{ap:C}{Driven-dissipative dynamics}
 
To compare the predicted spectrum with what would be observed in a
continuous-wave ESR-STM experiment, we model the dynamics of the
trimer using a Lindblad master equation restricted to the
low-energy manifold~\cite{Breuer2002}. This phenomenological approach captures the
steady-state response under coherent driving while accounting for
spin and chirality relaxation, as well as for finite-temperature
thermal occupation, through a small set of collapse operators.
 
\paragraph*{Driven Hamiltonian.}
The local AC magnetic perturbation acts on spin 1 only. It is given by
Eq.~\eqref{eq:driving} of the main text with $b(t) = b_0\cos(\omega t)$,
and is thus parametrized by an amplitude $\Omega = g\mu_B b_0/\hbar$, a
frequency $\omega$, and an angle $\alpha$ that controls the orientation of the
AC field relative to the static field $\vec B \parallel \hat z$. The full
Hamiltonian is $H(t) = H_0 + V(t)$, with $H_0$ the diagonal low-energy
Hamiltonian of Eq.~\eqref{eq:ham} and $S_1^x$, $S_1^z$ the projected
operators of Eqs.~\eqref{eq:Sz1_matrix} and \eqref{eq:Sx1_matrix}.
 
\paragraph*{Master equation.}
The reduced density matrix evolves according to
\begin{equation}
\dot\rho = -\frac{i}{\hbar}[H(t),\rho]
+ \sum_\mu \mathcal{D}[L_\mu]\rho,
\label{eq:lindblad}
\end{equation}
with the standard dissipator
$\mathcal{D}[L]\rho = L\rho L^\dagger - \tfrac12\{L^\dagger L,\rho\}$.
The collapse operators $L_\mu$ are chosen to model the physical
relaxation and dephasing channels relevant to the experimental
platform.
 
\paragraph*{Rate conventions.}
Since spin and chirality have their own relaxation and coherence times, and
since $\delta_\chi$ is comparable to $k_BT$ in the regime of interest, we first
define the rates entering Eq.~\eqref{eq:lindblad}. Each relaxation
channel exchanges population with the bath at a downward rate
$\Gamma_\downarrow$ and an upward rate $\Gamma_\uparrow$. Their ratio is fixed
by detailed balance,
\begin{equation}
\frac{\Gamma_{i\to j}}{\Gamma_{j\to i}}
= \exp\!\left(-\frac{E_j-E_i}{k_BT}\right),
\label{eq:detailed_balance}
\end{equation}
so that in the absence of driving the steady state of
Eq.~\eqref{eq:lindblad} approaches the Gibbs distribution
$\rho_{\rm th} \propto e^{-H_0/k_BT}$; their sum fixes the rate at which
equilibrium is restored, and we accordingly define
$1/T_1 \equiv \Gamma_\downarrow + \Gamma_\uparrow$. This is the quantity
accessed in pump-probe experiments, and the one for which
$1/T_2 = 1/(2T_1)+\gamma_\phi$ holds. 
For a pure-dephasing operator
$L = \sqrt{\gamma}\,A$, with $A$ Hermitian and diagonal in the eigenbasis of
$H_0$, populations are unaffected and the coherence between eigenstates with
eigenvalues $E_i$, $E_j$ decays at $\tfrac{\gamma}{2}(E_i-E_j)^2$; we normalize
$A$ so that each channel contributes exactly the $\gamma_\phi$ defined above.
 
\paragraph*{Dissipative channels.}
Four classes of dissipative processes are included.
(i)~\emph{Spin relaxation} connects states of opposite $S^z$ at fixed
chirality, generated by
$L^{(\tau)}_{\rm sp} = \sqrt{\Gamma_s^{\downarrow,\uparrow}}\,
|i_\tau^{-}\rangle\langle j_\tau^{+}|$ for each chirality sector $\tau=\pm$,
where the superscript indicates whether the state belongs to the lower or upper
Zeeman doublet, and $\Gamma_s^{\downarrow}+\Gamma_s^{\uparrow} = 1/T_{1,s}$.
(ii)~\emph{Chirality relaxation} flips the chirality at fixed spin
projection,
$L^{(\sigma)}_{\chi} = \sqrt{\Gamma_\chi^{\downarrow,\uparrow}}\,
|i_\sigma^{-}\rangle\langle j_\sigma^{+}|$ within each Zeeman sector, with
$\Gamma_\chi^{\downarrow}+\Gamma_\chi^{\uparrow} = 1/T_{1,\chi}$;
physically, this can arise from any dynamical term that breaks $C_3$
symmetry, such as stochastic fluctuations of the local field or 
phonon-induced fluctuations of the exchange interactions on the three bonds.
(iii)~\emph{Pure spin dephasing},
$L_{\phi,s} = \sqrt{2\gamma_{\phi,s}}\,S^z_{\rm tot}$, destroys
coherence between Zeeman sectors without changing populations and
contributes to the spin transverse relaxation rate through
$1/T_{2,s} = 1/(2T_{1,s}) + \gamma_{\phi,s}$.
(iv)~\emph{Pure chirality dephasing},
$L_{\phi,\chi} = \sqrt{\frac{\gamma_{\phi,\chi}}{2}}\,\tau_z$, plays the
analogous role for the chirality pseudospin. 

\AppSec{ap:D}{Direct chirality resonance: homodyne detection}

Within a single Zeeman sector the longitudinal local drive
$V(t) = \hbar\Omega\cos(\omega t)\,S_1^z$ projects to
$(\hbar\Omega/3)\cos(\omega t)\,\tau_x$ via Eq.~\eqref{eq:spinproj}, since with
$\cos\alpha = 0$ the two Zeeman doublets decouple exactly and the $\sigma_x$
piece of $S_1^x$ plays no role. This defines the effective two-level problem of
Eq.~\eqref{eq:Hchiral2level}, whose rotating-wave steady state is the Bloch solution of
Eq.~\eqref{eq:XY_app}, valid for $\Omega_R \ll \delta_\chi/\hbar$, with thermal
chirality polarization
\begin{equation}\label{eq:Z0}
  Z_0 = \langle \tau_z\rangle_{\rm eq} = -\tanh\!\left(\frac{\delta_\chi}{2k_BT}\right).
\end{equation}
The resulting lab-frame magnetization, Eq.~\eqref{eq:S1z}, oscillates at
$\omega$ about the equilibrium value $-1/6$, with no resonant DC component. Note
that only the longitudinal component is modulated: the projected $S_1^x$ is
off-diagonal in $\sigma$, so $\langle S_1^x\rangle = \langle S_1^y\rangle = 0$
within a Zeeman sector and the surface spin acquires no precessing transverse
moment.

This AC magnetization is nonetheless detectable through the homodyne mechanism
already demonstrated for singlet-triplet transitions in
ESR-STM~\cite{bae18,chen23}, and follows from the tunneling magnetoresistance
readout~\cite{chen23}. The tunnel current
\begin{equation}\label{eq:tunnelcurrent}
  I(t) = G_j\left[1 + a\,\langle\mathbf{S}_{\rm tip}\rangle\cdot\langle\mathbf{S}\rangle\right]
         \left[V_{\rm DC} + V_{\rm RF}\cos(\omega t + \phi)\right]
\end{equation}
contains the product of two quantities oscillating at $\omega$: the RF bias and
the local magnetization of the driven spin. Only their cross term survives the
time average, and using Eq.~\eqref{eq:S1z} together with
$\overline{\cos^2\omega t} = \overline{\sin^2\omega t} = 1/2$ and
$\overline{\cos\omega t\,\sin\omega t} = 0$ one obtains
\begin{equation}\label{eq:Ihom_app}
  \Delta I_{\rm hom} = \frac{G_j\,a\,S_{\rm tip}^{z}\,V_{\rm RF}}{6}
    \bigl(\langle\tau_x\rangle\cos\varphi + \langle\tau_y\rangle\sin\varphi\bigr),
\end{equation}
where the prefactor $1/6$ combines the factor $1/3$ of
Eq.~\eqref{eq:spinproj} with the $1/2$ of the time average, $S_{\rm tip}^{z}$ is
the tip polarization along the quantization axis, and $\varphi$ is the phase lag
between the RF bias that modulates the junction and the AC field that drives the
spin.

\end{document}